\documentclass{acm_proc_article-sp}

\usepackage{hyperref}

\renewcommand{\texttt}[1]{%
  \begingroup
  \ttfamily
  \begingroup\lccode`~=`/\lowercase{\endgroup\def~}{/\discretionary{}{}{}}%
  \begingroup\lccode`~=`[\lowercase{\endgroup\def~}{[\discretionary{}{}{}}%
  \begingroup\lccode`~=`.\lowercase{\endgroup\def~}{.\discretionary{}{}{}}%
  \catcode`/=\active\catcode`[=\active\catcode`.=\active
  \scantokens{#1\noexpand}%
  \endgroup
}

\begin{document}

\makeatletter
\let\origsection\section
\renewcommand\section{\@ifstar{\starsection}{\nostarsection}}

\newcommand\nostarsection[1]
{\sectionprelude\origsection{#1}\sectionpostlude}

\newcommand\starsection[1]
{\sectionprelude\origsection*{#1}\sectionpostlude}

\newcommand\sectionprelude{%
  \vspace{0em}
}

\newcommand\sectionpostlude{%
  \vspace{0.5em}
}

\let\origsubsection\subsection
\renewcommand\subsection{\@ifstar{\starsubsection}{\nostarsubsection}}

\newcommand\nostarsubsection[1]
{\subsectionprelude\origsubsection{#1}\subsectionpostlude}

\newcommand\starsubsection[1]
{\subsectionprelude\origsubsection*{#1}\subsectionpostlude}

\newcommand\subsectionprelude{%
  \vspace{0em}
}

\newcommand\subsectionpostlude{%
  \vspace{0.25em}
}
\makeatother

\title{Security and Privacy in Future Internet Architectures
}
\subtitle{Benefits and Challenges of Content Centric Networks}
%
%
%
%
%

\numberofauthors{1} 
%
\author{
%
%
\alignauthor
Roman Lutz\\
       \affaddr{College of Information and Computer Sciences}\\
       \affaddr{University of Massachusetts Amherst}\\
       \email{romanlutz@cs.umass.edu}
}

\maketitle
\begin{abstract}
As the shortcomings of our current Internet become more and more obvious, researchers have started creating alternative approaches for the Internet of the future. Their design goals are mainly content-orientation, security, support for mobility and cloud computing. The probably most popular architecture is called Content Centric Networking. Every communication is treated as a distribution of content and caches are used within the network to improve the effectiveness. While the performance gain of Content Centric Networks is undoubted, there are questions about security and especially privacy since it is not one of its main design principle. In this work, we compare the Content Centric Networking approach with the current Internet with respect to security and privacy. We analyze improvements that have been made and new problems that have yet to be resolved. The Internet of the future could be content-oriented, so it is essential to identify potential security and privacy issues that are inherent to the architecture early on.
\end{abstract}




\section{Introduction}
Since the beginnings of the Internet, the world has changed a lot. Back then, all computers were stationary and used mainly by scientists who trusted each other. The computers themselves had scarce resources and their Internet connections were very slow. The situation is mirrored in the paper by David Clark~\cite{clark1988design} from 1988 in which he describes the design philosophy of the current Internet. The top goals include fault tolerance, flexibility to allow different kinds of communications and inclusion of a variety of networks. Regardless of the extent to which these and other goals were achieved, we can conclude that the main design principles do not correspond to today's requirements. As a consequence, it is not surprising that we can observe a lack of security, privacy, support for mobility and efficiency among other shortcomings. Even though a lot of work has focused on adding these desired features, e.g. with encryption and new protocols to handle mobility, inherent problems of the architecture can only be overcome to a certain degree. For many years, researchers all over the world have tried to find new clean-slate architectures to replace the current Internet at some point in the future. Specifically, the NSF-funded projects Future Internet Design (FIND), Future Internet Architecture (FIA)~\cite{FIA} and FIA - Next Phase (FIA-NP) aim at designing the next-generation Internet. After a multitude of ongoing sub-projects in the early phases, FIA-NP includes only the three most promising remaining approaches: MobilityFirst~\cite{MFRutgers}, eXpressive Internet Architecture (XIA)~\cite{XIA} and Named Data Networking (NDN)~\cite{NDN}. \\
As the name suggests, MobilityFirst focuses on support for mobility at a time when the majority of devices are mobile. It uses a highly scalable Global Name System for name resolution. XIA aims at accommodating a variety of network services and being flexible enough for new innovations of the future. NDN is based on the Open Source project CCNx, which implements the idea of a Content Centric Network. Most communications on the Internet nowadays have the goal of retrieving content, and others can be modelled as an exchange of content. Since popular content is requested frequently, Content Centric Networks involve routers with integrated caches to improve the scalability. \\
All of the three projects state improved and inherent security as a primary objective. What they really mean is verifiable integrity of communications. Other security problems that already exist in the current Internet are not necessarily solved and there are even new security vulnerabilities. Apart from that, none of the research groups declared privacy as a main design objective. It is logical that this can have consequences for users.\\
In this paper, we focus on the content-oriented approach taken by NDN, the most popular project with the largest research community. We first describe the architecture of a content-oriented Internet in Section~\ref{sec:ccndesc}. Afterwards, we compare the current Internet with Content Centric Networks with respect to security and privacy. Section~\ref{sec:unchanged} describes what remains unchanged before Sections~\ref{sec:benefits} and \ref{sec:challenges} examine benefits and challenges of Content Centric Networking. Finally, Section~\ref{sec:related} gives an overview of related research followed by a conclusion.

\section{Content Centric Networking}
\label{sec:ccndesc}
\begin{figure}
\centering
\includegraphics[scale=0.25]{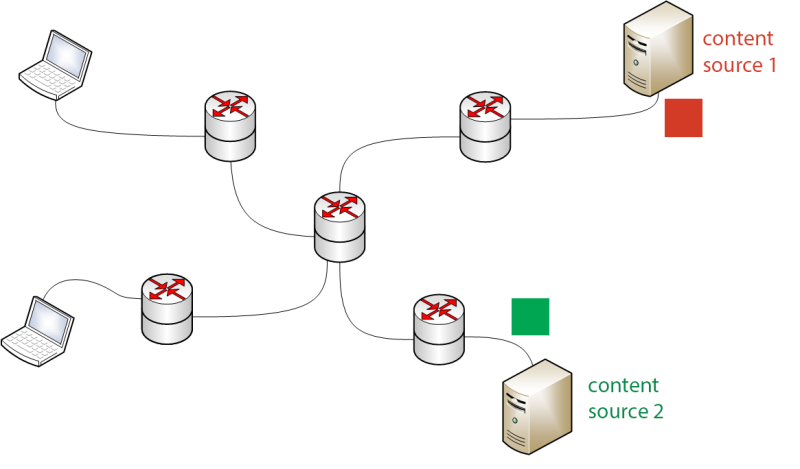}
\caption[]{An example of Content Centric Networking\footnotemark[1] (1). Two hosts are on the left, two content providers on the right. The routers in the network have caches.}
\label{fig:ccn1}
\end{figure}

\begin{figure}
\centering
\includegraphics[scale=0.25]{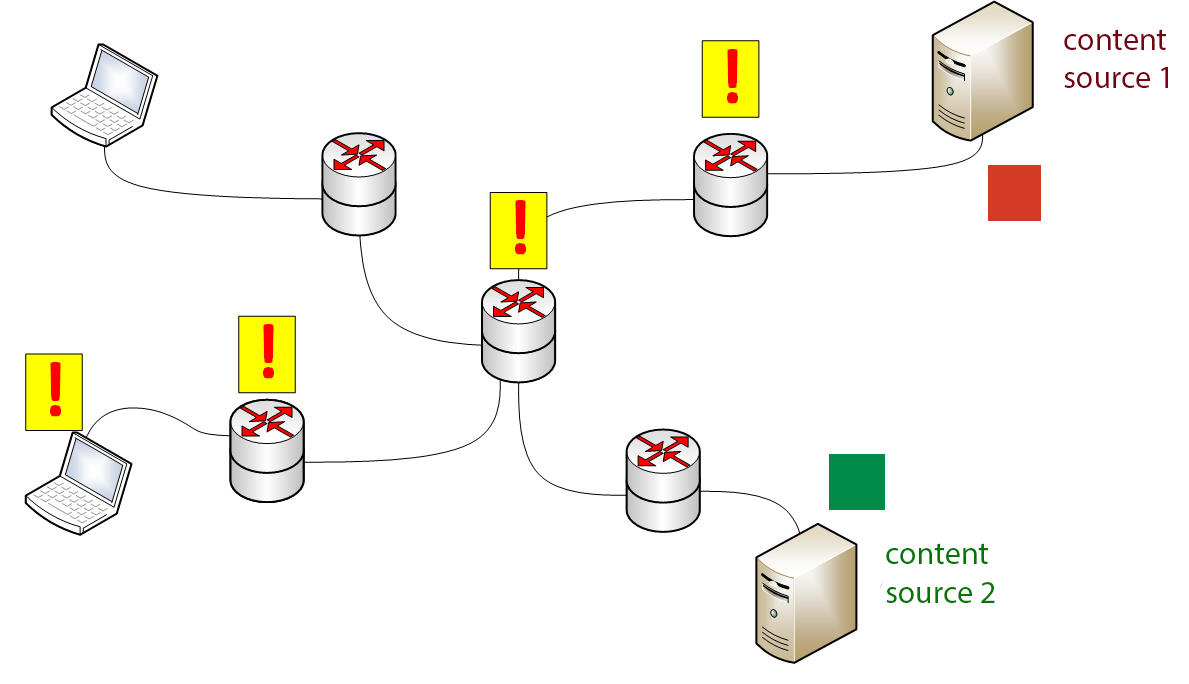}
\caption[]{An example of Content Centric Networking\footnotemark[1] (2). The host in the bottom left corner requested content from the content provider in the top right corner. None of the routers between them had the content cached, so the interest is forwarded.}
\label{fig:ccn2}
\end{figure}

\begin{figure}
\centering
\includegraphics[scale=0.25]{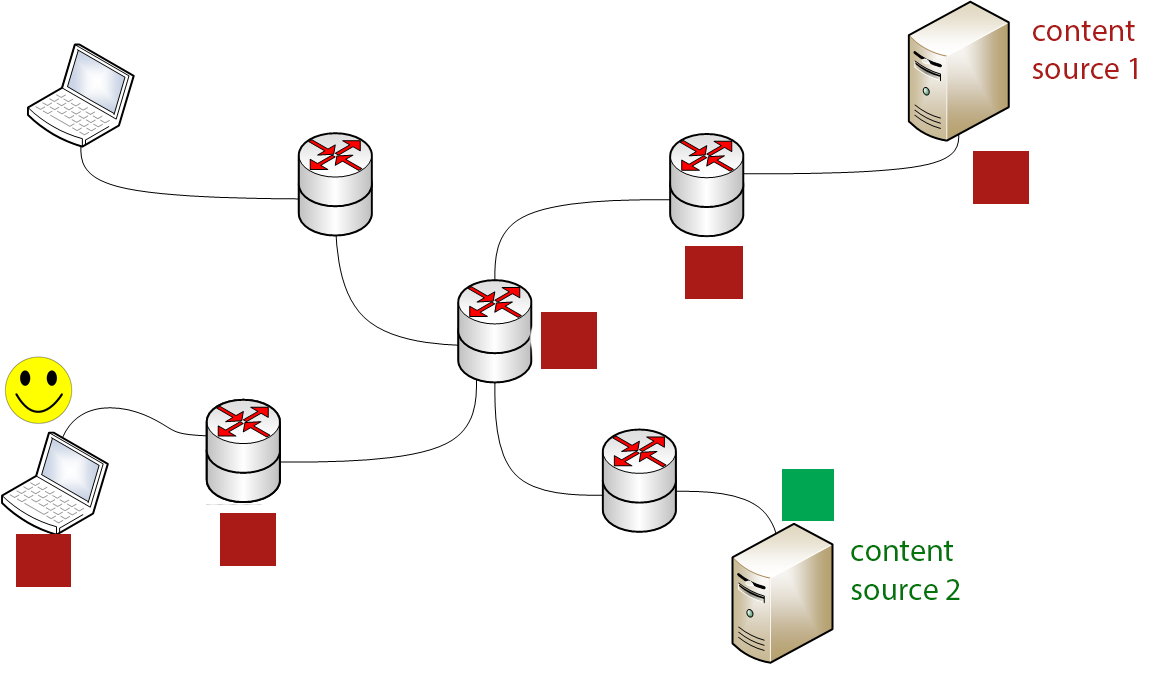}
\caption[]{An example of Content Centric Networking\footnotemark[1] (3). The content provider returned the requested content and name signed with his signature. The content is forwarded backwards in the same path the interest came. The routers can cache the content or not dependent on their cache replacement policy.}
\label{fig:ccn3}
\end{figure}

\begin{figure}
\centering
\includegraphics[scale=0.25]{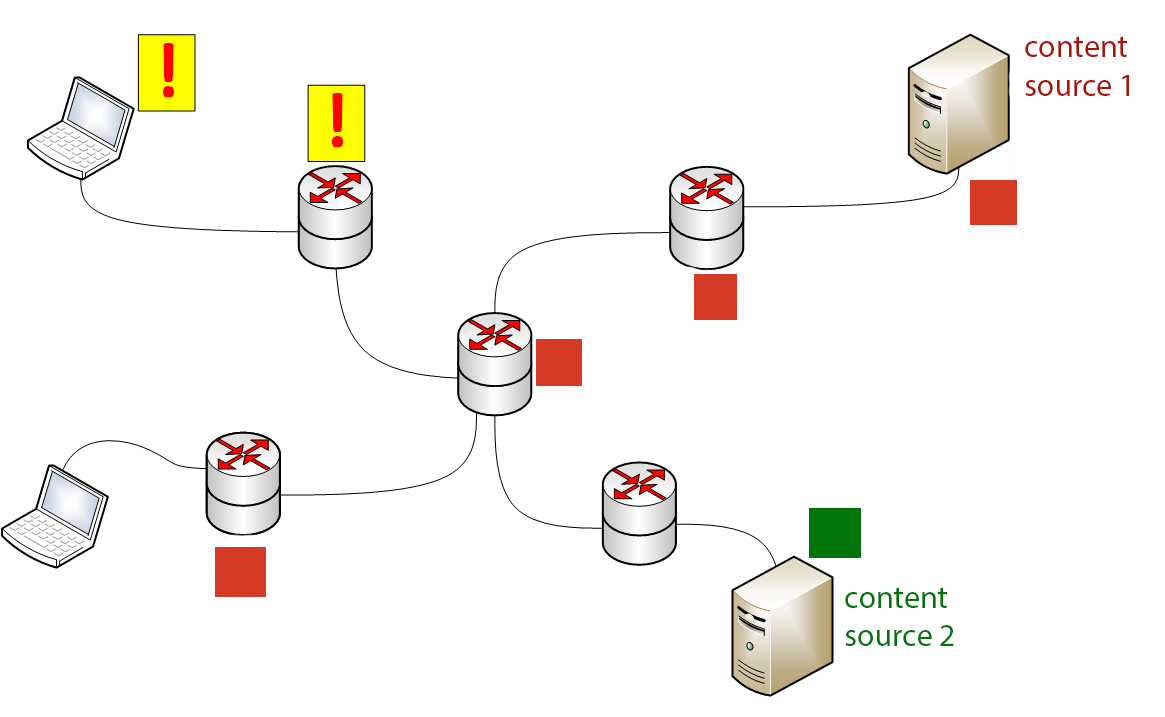}
\caption[]{An example of Content Centric Networking\footnotemark[1] (4). Another user requests the same piece of content that is already stored close by in a router's cache.}
\label{fig:ccn4}
\end{figure}

\begin{figure}
\centering
\includegraphics[scale=0.25]{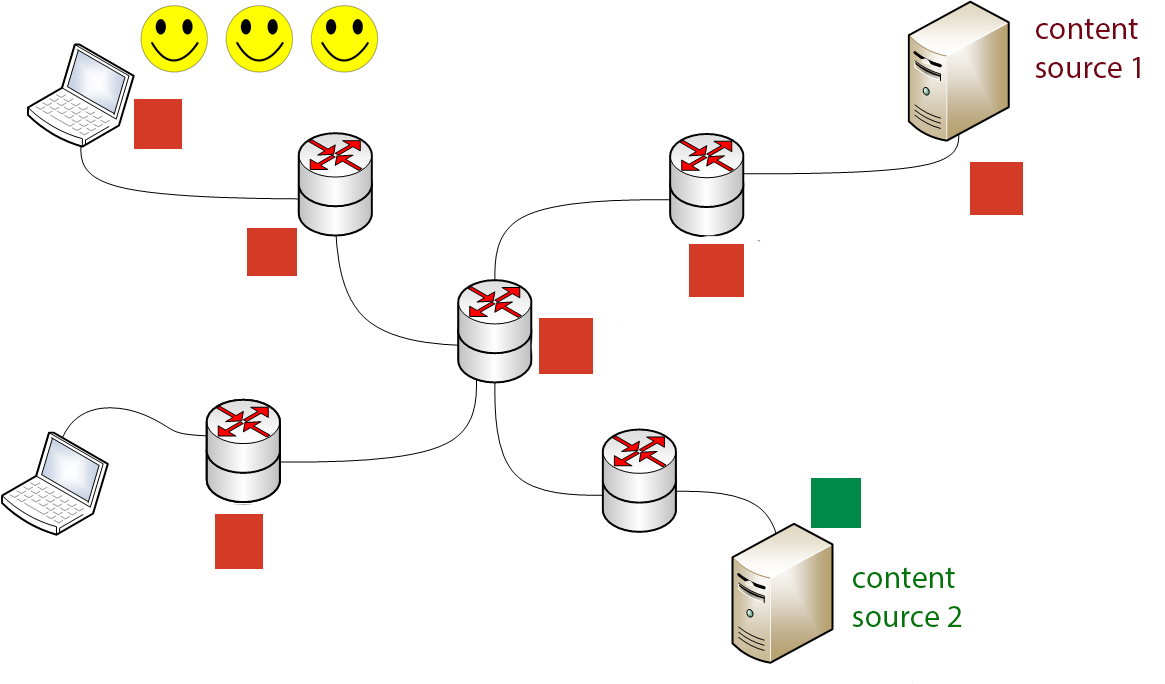}
\caption[]{An example of Content Centric Networking\footnotemark[1] (5). The router recognizes the content name as cached content and returns it.}
\label{fig:ccn5}
\end{figure}

The idea behind Content Centric Networking is to adapt the messages sent over the Internet to what they really are: content. Instead of of the restriction to end-to-end communications between pairs of users, Content Centric Networking allows for much more flexible and efficient message exchange. The main principle is that everything is content. In contrast to the current Internet's \textit{push} model, Content Centric Networking uses a \textit{pull} communication model. If a host wants to see a specific piece of content, he can simply request it by its name.  In both CCNx and NDN, content names are hierarchical, e.g. \texttt{/umass/cs/cs660/student/report}. In theory, flat naming could have been used instead, but that requires a name resolution. A discussion of benefits and problems of hierarchical and flat naming goes beyond the scope of this paper. The hierarchical content name is sent as a so-called \textit{interest} to the next router. Routers communicate with each other by sending name prefixes that they can serve. This is similar to routing based on IP prefixes in the current Internet. Before forwarding the interest to the next router, every router will first check whether the queried content is in its cache. If no router has the content in its cache, the interest is forwarded all the way until it arrives at the content provider. He will send the name of the content, the content itself and his signature back. The signature is just the hash of the name and content, \texttt{h(name, content)}, encrypted with the provider's private key. It allows the recipient to verify data and source integrity. For that, he needs to hash the name and content himself, decrypt the signature with the provider's public key and compare the them. This verification step is necessary, because hosts will often receive content from a cache instead of the content provider. An example scenario is shown in Figures~\ref{fig:ccn1}, \ref{fig:ccn2}, \ref{fig:ccn3}, \ref{fig:ccn4} and \ref{fig:ccn5}.\\
Compared to the current Internet, routers are very different. They basically contain three tables: the cache, the Pending Interest Table~(PIT) and the Forward Information Base~(FIB). Based on their cache replacement policy routers decide to cache content that is forwarded through them or not. The PIT registers every interest that the router receives and the interface from where it arrived. By looking up the name in the FIB, which maps name prefixes to outgoing interfaces, the router finds out where to forward interests if they are not cached. The way the FIB is used is very similar to longest prefix matching of IP prefixes in routers of the current Internet. Routers communicate with each other to fill and update the FIB. Apart from reducing the network congestion by returning cached content, routers also aggregate interests. If multiple interests for the same content arrive before the router receives the content, it only forwards the first interests and keeps all requesting interfaces in the PIT. Once the content arrives, it is forwarded to all of the requesting interfaces. This illustrates how multicast routing is possible without additional effort. In general, routers drop all interests from the PIT after the content is forwarded. If the content does not arrive due to loss, the interest will be dropped after a timeout. In Sections~\ref{sec:benefits} and \ref{sec:challenges}, we will describe the consequences this design has on security and privacy.\\
The content itself can be sent in the clear or encrypted. In order to allow for scalable communication, the names should not be encrypted at least for popular content in order to allow for the use of caching. If the names are encrypted, the requesting user will not know the name for which to query unless he knows the secret encryption key. We discuss this issue in greater detail in Section~\ref{sec:challenges}.\\

\footnotetext[1]{The pictures used for this graphic are from wikimedia.org and wikipedia.org. They are referenced at the end of this document.}

\section{Unchanged Security and Privacy compared to current Internet}
\label{sec:unchanged}
Even though the architecture is different from the current Internet, some problems have not been solved. Several of them have to do with censorship. Censorship Circumvention - or more generally privacy - is not defined as a primary design principle of Content Centric Networks, so additional techniques will have to be built on top of the architecture to support it. Firstly, the content provider is still easily identifiable. While it is possible to do this through his IP address in the current Internet, Content Centric Networking forces him to sign every piece of content with this signature. As a consequence, censors can check whether arriving content is coming from an unwanted source and possibly filter it. We will look into this problem in Section~\ref{sec:sigpriv}. Secondly, censors are currently able to block known Tor relays and proactively probe hosts in order to find out whether they are Tor bridges. If they suspect a host, they can simply behave as if they were a benign user and try to access Tor through the bridge. If it works, they have identified a Tor bridge. In Content Centric Networks, anonymity networks are also conceivable. Section~\ref{sec:privenhtech} examines existing solutions such as AND$\bar{a}$NA. Similarly to Tor, AND$\bar{a}$NA's relays can be blocked by censors. To our knowledge, bridges have not been proposed yet, but they could be tested similarly to proactive probing for Tor bridges. Thirdly, censors can filter interests for banned keywords or prefixes in names. Further techniques like Deep Packet Inspection (DPI) are also applicable. Finally, trusted third parties like Certification Authorities (CA) in the current Internet will still be necessary. Even though every piece of content has to be signed by the publisher, the recipient needs a trusted third party delivering the public key in order to be able to verify the integrity of the content.

\section{Security and Privacy Benefits of Content Centric Networks}
\label{sec:benefits}
In the following, we describe how the users' security and privacy benefit from the Content Centric Networking architecture. This includes verifiable integrity with signed content, the absence of addresses in interests, abandonment of name resolution in favor of hierarchical names and a certain degree of protection against Denial of Service (DoS) attacks. 
\subsection{Verifiable Integrity}
As described in Section~\ref{sec:ccndesc}, it is mandatory for content provi-ders to sign the content and name with their private key. Therefore, the recipient can verify the integrity of data and source. This prevents spoofing as it is possible in the current Internet. A positive side-effect of it is Routing Security. As every other piece of content, communication between routers has to be signed. This prevents adversaries from influencing the routing tables, i.e. the FIB in Content Centric Networks.
\subsection{Absence of Addresses}
Content Centric Networking abandons addresses totally. Neither interests nor the content delivery messages contain addresses. Interests in Content Centric Networks contain only the name of the requested content, but not who requested it. Only the first forwarding router knows the interface from which the content was requested. All other routers only know the previous router on the forwarding path. When the content provider returns the content, his message also includes the content name, his signature, the publisher's ID and information about where to retrieve the publisher's public key. As a result, there is no necessity for addresses, although the publisher can be inferred from the ID and the key. From a privacy perspective, this design improves anonymity because the source of an interest is unknown or at least hard to find out. In Section~\ref{sec:cacheprivacy}, we explain various techniques with which an adversary can still find out who sent the interest.\\
But the different communication paradigm has even more advantages. In the current Internet, many attacks require the attacker to directly send messages to the victim. In Content Centric Networks the victim has no address, so this threat is completely mitigated because no content can arrive at a host without the host requesting the content in advance. It is slightly different for content publishers. This case will be examined in the next section.
\subsection{Protection against Denial of Service}
\label{sec:protdos}
Apart from the protection against direct attacks on hosts, Content Centric Networking even offers a protection against Denial of Service (DoS) attacks on content providers. Assume an attacker launches a DoS attack on a publisher by sending lots of interests. These are simply collected at the first router and only one interest is forwarded. Even if the attacker controls a large botnet that is distributed over many locations that do not share routers on the routing path to the publisher, interest aggregation mitigates most of the attack. What actually happened is a shift in the point of attack from the content provider to the routers. Instead of flooding the publisher, the attack fills routers' Pending Interest Tables. We discuss this new DoS attack and possible coutermeasures in Section~\ref{sec:newdos}. Although neither NDN nor CCNx include this feature, some researchers have suggested adding a flag to interests that indicates that the interest should be routed to the content provider without returning cached content. Such a flag allows attackers to directly flood the publishers again.
\subsection{Name Resolution}
Content Centric Networks are based on hierarchical names and therefore do not require explicit name resolution. In the current Internet, name resolution is provided by DNS and offers a multitude of possibilities for attackers to exploit, e.g. DNS hijacking. If flat names are used instead, the advantage is lost. 

\section{Security and Privacy Challenges of Content Centric Networks}
\label{sec:challenges}
Even though there are security and privacy aspects that improve with Content Centric Networking, it also poses new challenges. In this section, we outline several of these concerns and suggest possible solutions. We start with Interest Flooding Attacks, the most studied attack on Content Centric Networks and go on to describe less known and privacy-related problems.
\subsection{Interest Flooding Attacks}
\label{sec:newdos}
As explained in Section~\ref{sec:protdos}, DoS attacks in Content Centric Networks will target routers instead of content providers. For that reason, they are called \textit{Interest Flooding Attacks} (IFA). They have been the subject of many research papers over the last few years. In this section, we describe the problem in general and a few simple countermeasures. For an overview of more sophisticated countermeasures, we refer to Section~\ref{sec:related}. To carry out an IFA, an attacker needs to sends lots of interests. If the interests are all for the same piece of content, the first router aggregates the interests in its PIT. As soon as the number of pending interests surpasses the PIT capacity, the router drops or rejects interests. The denial of service on the first router works at least until the content actually arrives at the router and the requests can be answered. Effectively, the attacker just degraded his own local network's performance by flooding the router. If a router closer to the core or close to a specific content provider is the target, the attacker adapts the attack to include various different interests for content by the same publisher. Using different content names helps against interest aggregation. Ideally, the attacker controls a botnet that is distributed over many places, such that interest aggregation does not take place too early. Some routers then become bottlenecks for interests, resulting in the same situation as before, albeit at different routers. The situation becomes even more troubling if the attacker controls a remote content source. He could make the source answer very slowly on purpose to delay the content, while at the same time filling up the PIT with interests.\\
There is a relatively simple countermeasure used in NDN. Routers could keep track of the number of interests per domain and drop some of them if there are too many in one specific domain. Against an IFA, this is an effective strategy, but also has some issues. If there is a very popular event, e.g. the Superbowl, that most people watch, and the router starts dropping the interests, this is actually counterproductive. Of course, this also applies to all kinds of popular content. In fact, it allows the attacker to selectively block certain content by sending lots of requests. Most of his interests are dropped, but the same applies to legitimate hosts.\\
Depending on the configuration of the routers, they might verify the integrity of the content themselves. This involves expensive operations that can slow down the router. Especially in a scenario with a botnet and a cooperating content provider, the routers' speed could be reduced. In order to handle high load situations, routers could stop verification as soon as it becomes a burden for their forwarding speed.\\
Similarly, an adversary could try to degrade the network performance by filling the cache with arbitrary data. This is called \textit{cache pollution}. Again, this works best with a cooperating content provider delivering the content and a botnet requesting it. The attack basically works with any kind of content that is not of any use to other hosts. Such a scenario is detectable by comparing cache hit rates. If only the same people request the content over and over again and nobody else is interested in it, it is very likely a cache pollution attack. It is more difficult to defend against such an attack. Countermeasures include blacklisting the cooperating source provider such that its content is not cached any more and ignoring the interests of recognized bots for caching or even completely.

\subsection{Cache Privacy}
\label{sec:cacheprivacy}
The main privacy concern in Content Centric Networks is caching. Having a copy of my communications stored in caches that are available to everybody who knows how to query it is a risk. In the current Internet, adversaries have to sniff on traffic exactly at the time it is sent, because there are supposed to be no traces of the actual content. This is substantially harder than being allowed to query the messages by name and even some time after they were sent. As a consequence, the use of caches causes a tradeoff between privacy and efficiency. In the following, we will examine attacks and possible countermeasures.\\
There are basically three attack scenarios as identified by Lauinger~\cite{lauinger2010security} and Acs et al.~\cite{acs2013cache}:
\begin{itemize}
\item Cache Enumeration
\item Timing Attack 
\item Conversation Cloning
\end{itemize}
For each of them, an adversary simply needs to be in the same local network as the victim such that they share a cache. None of the attacks actually involve illegally sniffing on other people's traffic. Instead, attackers can simply request items from the caches like any legitimate user.
\subsubsection{Cache Enumeration}
There are mainly two reasons why someone might want to enumerate a cache. If the cache population, i.e. the people who share a cache, is small, it could be possible to infer who has requested specific content just by looking at all cached content. In many cases, the name of the content will involve some user-specific part, e.g. \texttt{/provider/mail/username}, that immediately gives away the identity of the requesting person. In addition to that, the signature of the content source could be enough to infer the communication partner. For censors, it might even be sufficient to find out that \textit{someone} queried specific content. Otherwise, finding out which content is in the cache is a first step before a Timing Attack or cloning a conversation.\\
The feasibility of Cache Enumeration arises from routers allowing users to either directly request all cached data, e.g. with a command like \texttt{ccnls} in CCNx, or by using a combination of prefix matching and the exclude functionality of NDN as explained by Lauinger et al.~\cite{lauinger2012privacy2}. Prefix matching is performed on every query, so if the cache includes \texttt{/email/work/2015} and \texttt{/email/private/2015} and the adversary requests \texttt{/}, an arbitrary cached content is returned because all entries match. Then, the exclude functionality is used to query again excluding the first content name. While this may take lots of requests for a large cache, the distance to the cache is very small. Still, it is unlikely that the attacker will get a consistent view of the cache.
\subsubsection{Timing Attack}
Timing Attacks aim at finding out if and when a user requested a specific piece of content. The adversary can measure the time it takes for the content to arrive and compare it to previously measured times to find out whether his request resulted in a cache hit or not. Therefore, measuring round-trip-times is a preliminary step for a Timing Attack. It also requires the attacker to know in advance which content the user might query. Censors, for example, could have a number of blacklisted files and check whether a user requested them. But even local adversaries with limited resources can execute such an attack to spy on people within the same cache population. Lauinger et al.~\cite{lauinger2012privacy2} describe how Timing Attacks can be carried out invasively and non-invasively. Here, \textit{invasive} means that the cache treats a request as any other request and possibly updates its state, while a non-invasive request would not influence the cache's state and simply retrieve the cache's content. Which method is chosen depends solely on the cache itself and whether it allows non-invasive queries. Disallowing them makes it harder for the adversary, although not impossible. In the invasive scenario, the attack depends on the cache replacement policy. For example in the case of FIFO (First In First Out), LRU (Least Recently Used) or Random Replacement, the attacker first needs to find out the \textit{cache lifetime} or \textit{characteristic time} $t_c$ of the cache, i.e. the expected time for which the cache stores a piece of content. For LRU, it is the time a file remains in the cache while not being queried. The attacker can then repeatedly request the content every $t_c + \epsilon$ time units for a small $\epsilon$. If he gets a cache hit, someone within the cache population queried the content within the last interval. Especially if a user-specific content name is used, the attacker knows who requested the content. Since the characteristic time is determined as an expected value, the attacker will get a result that is true only with a certain confidence. Another problem for the adversary is that the targeted user could request the content shortly after the attacker, i.e. within $\epsilon$ time units, and the request would therefore not be detected. CCN offers a way to overcome this, though: Every cached content is stored as chunks of $4$KB which can individually be requested. The attackers can thus iterate over the chunks of the given content and query one of them every $t_c$ time units. Lauinger et al.~\cite{lauinger2012privacy2} call this Parallel Cache Probing. It is important to notice that these methods are dependent on the cache replacement strategy. With a more complex method using randomness it could become infeasible for a local adversary to launch a Timing Attack. For example, in order to understand which content is cached, one might have to observe all requests at a router. This could restrict this attack to more powerful adversaries like a censoring country which might have access to the router anyway and will not use this method.
\subsubsection{Conversation Cloning}
The idea behind Conversation Cloning is that an attacker tries to get the whole content corresponding to a data flow. This flow could, for example, be a video call using Voice-over-CCN (VoCCN)~\cite{jacobson2009voccn}. The content itself is assumed to be encrypted. But by reassembling the messages of the conversation, the adversary could potentially find a side channel and infer privacy-related information about the content. Possible side channels include the size of the messages and their timings. A similar attack is possible on VoIP~\cite{wright2008spot}, so it is likely to be a threat for VoCCN, too.\\
Somehow, adversaries need to find out about the names that are used in the data flow. It is likely to have some kind of serial number, e.g. \texttt{/voccn/call/alice/1}, followed by \texttt{/voccn/call/alice/2} and so on. With the previously discussed Cache Enumeration, the attacker can easily find out about a current sequence number and possibly the naming scheme. This enables him to predict names of future messages and to request and receive the messages in real-time like the legitimate receiver. As discussed in the next section, the naming issue can be overcome by encrypting parts of the names or even the whole name. If the attacker is not able to retrieve the content in the right order, he is at a significant disadvantage, because he can not infer the order from the content itself as it is encrypted. The main problem is still that Cache Enumeration is possible in the first place. Using the exclude functionality, every message with the known routing prefix \texttt{/voccn/call/alice/} can be queried by attacker.
\subsubsection{Countermeasures}
This section contains a list of possible countermeasures against the previously described attacks. They include - but are not limited to - the work of Lauinger~\cite{lauinger2010security}, Lauinger et al.~\cite{lauinger2012privacy1, lauinger2012privacy2}, Chaabane et al.~\cite{chaabane2013privacy}, Acs et al.~\cite{acs2013cache}. Most of them affect the efficiency. As a result, we get a trade-off between privacy and efficiency.
\begin{enumerate}
\item Disallow caching in general. This eliminates one of the main benefits of Content Centric Networks.
\item Disallow non-invasive queries, i.e. a query should always affect the cache's state and not just allow users to enumerate the contents without changing the state.
\item Disallow techniques that are used for Cache Enumeration, such as \texttt{ccnls}. In the context of DNS Snooping such requests are often called non-recursive.
\item Restrict or completely disallow the exclude functionality of caches.
\item Disallow requesting chunks of a piece of content to prevent Parallel Cache Probing.
\item Encrypt names in interactive conversations. Encryption prevents the adversary from predicting names. The content is not meant for people who do not know about the naming scheme anyway. Therefore, the efficiency of the network is not degraded.  
\item Tunnel traffic using an anonymity system. This has basically the same effect on Cache Privacy as Encryption. AND$\bar{a}$NA~\cite{dibenedetto2011andana} is an example of such an anonymity system.
\item Use a more complex cache replacement policy. The Timing Attacks rely on knowledge about the cache lifetime. If the cache uses a complex strategy, the attacker has a harder time understanding it. Attacks based on the cache lifetime might not be possible at all any more.
\item Only cache popular content. Popular content is requested often and by multiple people. This can be enforced with a cache replacement policy that adapts to the content that is forwarded by the router. Current research is focusing on such topics.
\item Make cache lifetimes very short. This limits the time for adversaries to retrieve the requested content, but at the same time affects the performance.
\item Choose cache lifetime randomly for each cached file to prevent attackers from measuring the cache's characteristic time.
\item Increase the anonymity set, i.e. the cache population. This results in more people being served by the same cache and thus in reduced efficiency.
\item Add a minimum response delay in case of a cache hit. The actual delay could be chosen randomly in order to prevent attackers from measuring the minimum response delay. It obviously increases the total delay, but if chosen carefully the adversary can not distinguish whether the content was cached in the closest cache or several hops away.
\item Do not cache private content. This is similar to only caching popular content, but instead identifying content that should not be cached. The key observation is that private content should only ever be received by the communicating parties and caching it will neither improve their privacy nor the overall performance of the network, because nobody else can possibly legitimately want the content. Based on this, Lauinger et al.~\cite{lauinger2012privacy1, lauinger2012privacy2} propose selective countermeasures, i.e. countermeasures that are only applied to privacy-related content. One way to implement this is by using a do-not-cache flag in interests. But privacy-aware users might decide to use it on all their interests, so it might be better to have the content origin set the do-not-cache flag. The disadvantage of this option is that the receiver has no choice. Furthermore, caches of routers on the return path could simply decide to ignore the flag and still cache the content.
\end{enumerate}
Apart from these countermeasures, it might already be enough to detect attacks on Cache Privacy. If an attacker is identified, his requests could be ignored partially or completely to resolve the situation. The detection is unfortunately not easy and prone to finding false positives. The detection methods include the following:
\begin{enumerate}
\item Edge routers keep track track of how often the same content is queried from a single interface. If periodic querying occurs, it is assumed to be a Timing Attack or Cache Enumeration.
\item Edge routers remember the recent hit rate for each interface. If an unusually high hit rate is detected, it is likely because of Cache Enumeration.
\item Edge routers keep track of requests using the exclude functionality. As a consequence, they can find out if somebody is trying to clone the conversation.
\end{enumerate}
All these methods should be deployed as close to the users as possible. Still, all of them are suboptimal. Firstly, periodic querying might be necessary in a company that wants to know the state of some content. For example, a news agency always has to know whether it is missing a story that another agency has already put on their website. Secondly, a high hit rate could be accidental. For example, if multiple people are watching the same movie, but they did not start exactly at the same time, there will be lots of cache hits resulting in false positives. Thirdly, the exclude functionality can be used in a legitimate way. If a user makes extensive use of it, he could falsely be detected as an attacker.\\

\subsection{Name Privacy}
With a content-oriented architecture, the performance benefits come from having a publicly known and often human-readable name for the content and thus being able to match requested content names to cached content and to aggregate interests. But users sacrifice their privacy for using these names. Assume that a user lives in a censoring country and wants to look at blocked content. Retrieving the content with its human-readable name is impossible, because the censors can filter out unwanted names. If the content were already cached at the time the censors decide to block it, they could simply delete content with the given name from all caches. A powerful adversary could even store interests to analyze them later. Caches allow censors to find unwanted content even after it is sent. Even for privacy-aware users that do not want anybody to track their interests it is not desirable to send names in the clear. As a result, Name Privacy is a challenge in Content Centric Networking.\\
There are conceivable countermeasures, some of which are inspired by techniques used in the current Internet.
\begin{enumerate}
\item Encrypt content names. If only the last part of the name is encrypted, e.g. \texttt{/provider'/mail'/username'/a2e13f7b5}, there is still a fairly good chance to guess what the content is. Encrypting the whole name would result in routing problems because the hierarchical structure is not used any more. Instead, a part of the name should be encrypted, e.g. \texttt{/provider'/mail'/c298a67fe'/a2e13f7b5}. The encryption can not be extended over the name of the provider, because the interest would not arrive there due to a non-matching name prefix. In many cases, though, censors will completely block a provider. We conclude that the hierarchy of names helps censors distinguish between benign and unwanted content. As a consequence, encryption of names is only a viable option for increased privacy, but not sufficient for censorship circumvention. It is worth noting that the encryption has to be negotiated before the interest and that advantages of Content Centric Networks such as caching are lost.
\item Decoy Routing. Decoy Routing is essentially redirecting traffic once it is out of the reach of a censor. As effective as it is, Decoy Routing would require architectural changes. For a comprehensive description and analysis, we refer to \cite{karlin2011decoy, houmansadr2014no}.
\item Change names. To circumvent censors' blacklist of names, one could simply change the name in order to make it seem to the censor as if the content is normal and should not be filtered out. The problem is that the censors are likely to get suspicious if a content provider that is known for producing unwanted content suddenly offers lots of unblocked content. To check the assumption, they can query the content themselves. Furthermore, suspicious content providers are likely to be completely blocked by censors anyway. Therefore, changing names is at least not a permanent solution to censorship circumvention.
\item Ephemeral names. The idea is to generate new names for blocked content. Users should be able to predict the names while censors should not. This could maybe be achieved by having a separate generation algorithm for every user. In any case, with ephemeral names the benefits of the content-oriented architecture are lost.
\item Anonymity Systems. Relaying encrypted traffic through multiple other nodes is certainly an option, although again the advantages of Content Centric Networking are not used. An anonymity system similar to Tor in the current Internet is AND$\bar{a}$NA. We describe AND$\bar{a}$NA in Section~\ref{sec:andana}
\end{enumerate}
All in all, there is no perfect solution so far. This will surely be addressed by researchers.
\subsection{Signature Privacy}
\label{sec:sigpriv}
While we previously stated that integrity verification is a benefit of Content Centric Networking, it also provides problems. Content providers are identifiable by their signature, so censors can block content if they recognize an unwanted publisher. To mitigate this risk, we suggest two countermeasures.
\begin{enumerate}
\item \textit{Group} or \textit{ring signatures}. A number of content providers share a group signature. This can be administrated by a group manager (group signatures) or through interaction of the group members (ring signature). If a publisher whose content is filtered out by censors uses a group signature that he shares with a number of unblocked and popular publishers, there is a chance the censors will not block his content any more. Otherwise, the censors risk losing the services of the other popular publishers as well. It is unclear, though, whether popular services would take the risk of getting blocked themselves without incentives. There has been lots of research on group signatures that we recommend for more information, specifically \cite{chaum1991group,boneh2004short,bellare2003foundations,camenisch1997efficient}.
\item Ephemeral identities. A content provider could generate ephemeral identities for every piece of content. With a signature that is unknown to the censor, the content could go unblocked. Even if the censor decides to check the origin of the content, it would at least affect the performance of the censors network and add expensive computations. Whether or not censors are willing to do this remains to be seen.
\item Proxies. Instead of ephemeral identities, proxies could be used. The signature then becomes the signature of the proxy. If a proxy is used for unwanted content frequently, censor will eventually block the proxy's forwarded content.
\end{enumerate}
All the proposed solutions seem promising. The effects are all speculative, though, until they are actually used in a real-world scenario.
\subsection{Content Privacy / Access Control}
As in the current Internet, access control is enforced by content providers through encryption. This becomes even more important since their content might not even be coming directly from the publisher, but from a cache. If the encryption is specific to one user, the effect of caching is nullified. Therefore, Fiat et al.~\cite{fiat1994broadcast} propose broadcast encryption which allows $n$ users with different keys to decrypt the same encrypted content. While this allows effective caching, this method produces keys of length $\mathcal{O}(\sqrt{n})$ and increases the length of the encrypted content by a factor $\mathcal{O}(\sqrt{n})$. A different solution is atomic proxy cryptography offered by Blaze et al~\cite{blaze1998divertible}. The content provider encrypts content such that the first user can decrypt it and provides a proxy with a re-encryption key. The proxy has no knowledge about the content and simply applies the key and delivers the re-encrypted content to the second user, re-encrypts for the third user etc. The main problem of atomic proxy cryptography is the use of asymmetric keys that require computationally expensive operations. 
\subsection{Accountability}
The previously claimed advantage of having no addresses can be a problem. Law enforcement, for example, has to be able to trace back who requested illegal content. There are basically two outcomes: Either law enforcement is given access to the whole infrastructure such that they can keep track of interests in illegal content, or not. If not, users can not be held accountable for requesting illegal content. This aspect of Content Centric Networking has barely been studied, but is of great importance from a legal perspective.
\subsection{Revocation and Removal}
There are many reasons why somebody would like content to be remove content from caches:
\begin{itemize}
\item Revocation. Content can become outdated, especially if hosts repeatedly request the same piece of content and it has been in the cache for a long time. Content providers need a mechanism to either update or at least revoke outdated content.
\item \textit{Content Poisoning}. Fake content is in the cache and should be removed. Ghali et al.~\cite{ghali2014needle} examine this in detail. Basically, content poisoning occurs when the signature is invalid or not verifiable. An adversary can try to distribute the fake copy of content. This problem can easily be mitigated by having routers verify signatures, but - as discussed earlier - this causes the routers to perform expensive computations.
\item \textit{Cache Pollution}. This attack is discussed in Section~\ref{sec:newdos}. An attacker tries to fill the cache with arbitrary data to degrade the network's performance.
\item Illegal content removal. Criminals might try to keep illegal data in a cache by repeatedly querying it. The availability of illegal data within the network infrastructure is not acceptable, so there should be a removal mechanism.
\end{itemize}
While there are more sophisticated countermeasures, we restrict the following list to general purpose measures that achieve the content removal:
\begin{enumerate}
\item Explicit Removal. The service provider owning the cache could manually access and remove content. Assuming a large number of such cases, this method is infeasible.
\item Blacklist broadcasting. If the detection of unwanted content is automated, a blacklist of content names could be created that is broadcasted among routers. If a router receives the list, it deletes all cached content that is present in the blacklist. The broadcast messages cause a large communication overhead.
\item Periodic revalidation. Routers periodically verify that the cached content is still up-to-date. Like the previous approaches, periodic revalidation imposes a communication overhead for the routers. 
\end{enumerate}
There is certainly more research to be done on content removal from caches, because the described methods all have issues. Previous research has mainly aimed at detecting unwanted content before it is stored at all, but there need to be mechanisms to remove content that is not detected or identified as illegal later.

\subsection{Privacy Enhancing Technologies for Content Centric Networking}
\label{sec:privenhtech}
There are many \textit{Privacy Enhancing Technologies} (PET) for current Internet, but it has yet to be checked which are applicable to Content Centric Networking. Also, there might be new techniques conceivable. In this section, we present AND$\bar{a}$NA, which is to our knowledge the only currently existing anonymizing system for NDN.
\subsection{ANDaNA}
\label{sec:andana}
AND$\bar{a}$NA~\cite{dibenedetto2011andana} is an attempt to build Tor~\cite{dingledine2004tor} for NDN. It is designed as an overlay network on top of the NDN architecture. Important elements of Tor were used, such as \textit{Onion Routing}. The goal of AND$\bar{a}$NA is to provide anonymity through layered encryption. The assumption is that there is no adversary with the power of global surveillance. Under that assumption, traffic is relayed twice through so-called \textit{anonymizing routers} (AR) which are hosts or routers distributed all over the world. Each AR adds another layer of encryption. Compared to Tor, only two relays are necessary to achieve the same degree of anonymity. This is due to the absence of addresses in Content Centric Networking. The first step in a communication through AND$\bar{a}$NA is building an ephemeral circuit. It is used for only one interest and closed afterwards. For that, the user distributes a separate symmetric key to each of the ARs which are chosen from a public list in advance. Once the ephemeral circuit is set up, the interest is sent through the tunnels. The last part from the exit AR to the content provider is the actual communication as it is normally performed. After receiving the content, the exit router encrypts the content, original name and signature with the symmetric key. This ciphertext is the content that he forwards to the entry AR. The content and name are signed by the exit AR. The same happens at the entry AR. When the user finally receives the content, he has to decrypt it with the symmetric keys and verify the producer's signature.\\
The crucial point about anonymizing systems is that their delays should be relatively small in order to still provide a usable service. In their evaluation, DiBenedetto et al.~\cite{dibenedetto2011andana} show that AND$\bar{a}$NA outperforms Tor in the current Internet for small files up to $10$MB. Furthermore, the caching benefits of Content Centric Networks are lost because of the encryption. AND$\bar{a}$NA does not provide solutions for most known problems of Tor such as blocked relays and the distribution of a list of relays.

\section{Related Work}
\label{sec:related}
Chaabane et al.~\cite{chaabane2013privacy} thoroughly describe challenges to privacy in Content Centric Networks. They divide privacy threats into Cache, Content, Name and Signature Privacy and discuss possible countermeasures. Their work provides a great overview and starting point for further research on the topic.\\
Lauinger~\cite{lauinger2010security} looks mainly at DoS- and Cache Privacy-related problems of Content Centric Networking. His detailed analysis not only shows the weaknesses of the approach, but he also provides a comprehensive discussion of potential solutions.\\
Acs et al.~\cite{acs2013cache} examine Cache Privacy for NDN in great detail. They show how easy adversaries can find out about previously queried content and thus reconstruct a user's communication. In order to overcome this issue, they propose several countermeasures like delays and explicit privacy bits. Similarly, Lauinger et al.~\cite{lauinger2012privacy2} look at the effect of different caching strategies on Cache Privacy. They also describe how caches can enumerated by simply using functionality provided by CCN. \\
Arianfar et al.~\cite{arianfar2011preserving} consider Content Centric Networks as a step backwards in terms of privacy. To counter this, they suggest mixing a censored file with a cover file to achieve computational asymmetry. While this method does not provide guarantees, it increases the effort censors have to invest.\\
DiBenedetto et al.~\cite{dibenedetto2011andana} propose AND$\bar{a}$NA, an onion routing network designed as an overlay network to NDN. Similarly to Tor, AND$\bar{a}$NA aims at providing anoymity to users by relaying traffic and with layered encryption.\\
Ghali et al.~\cite{ghali2014elements} discuss content poisoning extensively. In order to prevent this problem, they argue that content verification should be left to the communicating parties instead of the routers. They propose small changes in the communications of NDN in order to establish trust management on the network layer.\\
Xie et al.~\cite{xie2012enhancing} propose CacheShield, a robust cache scheme applicable to CCN. CacheShield tries to optimize performance by caching only popular content. Their studies show that CacheShield also helps to mitigate cache pollution attacks. Teoli~\cite{teoli2013cache} evaluated CacheShield more extensively and confirmed its effectiveness. Conti et al.~\cite{conti2013lightweight} argue that CacheShield requires too much storage space. As an alternative, they propose a lightweight mechanism for the detection of cache pollution.\\
Ghali et al.~\cite{ghali2014elements} describe content poisoning in detail and come up with a solution that involves trust management on the network layer. Effectively, the introduce a so-called Interest-Key-Binding (IKB) which happens when sending the interest.\\
A lot of research groups have focused on the detection and mitigation of Denial of Service attacks in Content Centric Networks or Interest Flooding Attacks (IFA). W\"ahlisch et al. \cite{wahlisch2013backscatter}
analyze the problem space of attacking a CCN backbone, including an
experimental study of different IFA scenarios in NDN. 
Gasti et al.~\cite{gasti2013and} broadly discuss DoS and DDoS attacks in NDN and suggest a number of mitigation techniques. Compagno et al.~\cite{compagno2012ndn} experimented with a countermeasure based on router statistics. They show that even under attack, they can achieve a forwarding percentage of more than $80\%$ of legitimate traffic.
Ding et al.~\cite{ding2014cooperative} detect IFAs by using an entropy-based model. Dai et al.~\cite{dai2013mitigate} suggest using tracebacks in case of a suspected attack. By using the PITs of routers, they manage to find the origins of the attack and limit their access to the network by dropping interests. Wang et al.~\cite{wang2014detecting} describe how attackers could circumvent interest aggregation by requesting non-existent content. As a countermeasure, they suggest a threshold-based detection and mitigation scheme (TDM) that recognizes when too many interests time out. As an alternative, Wang et al.~\cite{wang2013decoupling} propose Disabling PIT Exhaustion (DPE) to reduce the effect of IFAs by directly recognizing malicious requests before the PIT is full. Afanasyev et al.\cite{afanasyev2013interest} performed large-scale simulations to test the effectiveness of several mitigation algorithms. Especially the satisfaction-based pushback algorithm, which enforces a limit on the number of forwarded interests based on an interface's interest satisfaction ratio, achieved promising results. Li et al.~\cite{li2014interest} propose Interest Cash, a countermeasure against IFA that mandates solving a puzzle before being allowed to send interests. Their evaluations show that the an attacker needs more than $300$ times more resources than a publisher to launch a successful IFA against him. Al-Sheikh et al. \cite{aws-rcani-15} compare nine different
countermeasures and discuss their limitations in different network
scenarios.\\

\section{Conclusion}
In this paper, we provide an overview of how various aspects of security and privacy change with Content Centric Networks. Even though security is a main design principle, the architecture allows for new attacks such as Interest Flooding Attacks, Cache Pollution and Content Poisoning. Furthermore, privacy is not taken into account which leads to new problems categorized as Cache, Content, Name and Signature Privacy. Overall, Content Centric Networking is intended to be a scalable and efficient architecture. It is neither completely beneficial nor completely detrimental for security and privacy. Instead, some aspects were improved while others offer new vulnerabilities. In general, we conclude that there is a tradeoff between the effectiveness of caches and privacy. This implies a change of methods for both attackers and defenders. In the current Internet, security was mainly built on top of the architecture. It remains to be seen whether more security and privacy can be added to Content Centric Networks. Considering that actual projects like CCNx and NDN are still being improved and changed, it is vital for researchers to find flaws and solutions in the architecture now while changes are still possible.

\section{Acknowledgments}
Thanks to Amir Houmansadr for stimulating discussions in several meetings throughout the semester.

\section{References}
%
\nocite{*}
\bibliographystyle{abbrv}
\bibliography{Bibliography}  

\end{document}